\documentclass[twocolumn,superscriptaddress,longbibliography,prb,preprint numbers,floats]{revtex4-1}
\usepackage{bm}
\usepackage{graphicx}
\usepackage{amssymb}
\usepackage{amsfonts}
\usepackage{amsmath}
\usepackage{epstopdf}
\usepackage{ulem}
\usepackage{color}
\usepackage{lineno}
\usepackage{lipsum}
\usepackage[urlcolor=blue,colorlinks=true,citecolor=blue,linkcolor=blue,p
dfstartview={FitH},bookmarks=false]{hyperref}
\usepackage{verbatim}

\usepackage{fancyhdr}
\usepackage{multirow}
\usepackage{array,mathtools,amssymb}
\newcolumntype{C}{>{$}c<{$}}

\definecolor{mygreen}{rgb}{0,0.65,0}%

\def \IFPAN{Institute of Physics, Polish Academy of Sciences, Aleja 
Lotnikow 32/46, PL-02668 Warsaw, Poland.}
\def \Wurtzburg{Physikalisches Institut (EP3), Universit\"at W\"urzburg, 
97074 W\"urzburg, Germany.}
\def \ITI{Institute for Topological Insulators,  Universit\"at 
W\"urzburg, 97074 W\"urzburg, Germany.}
\def \CSIS{Center for Science and Innovation in Spintronics, Tohoku 
University, Sendai 980-8577, Japan}

\begin{document}
\title[CuMnSb]
{Coexistence  of Antiferromagnetic Cubic and Ferromagnetic Tetragonal Polymorphs in Epitaxial CuMnSb}

\author{A. Ciechan}\affiliation{\IFPAN} 
\author{P. D{\l}u{\.z}ewski}\affiliation{\IFPAN} 
\author{S. Kret}\affiliation{\IFPAN} 
\author{K. Gas}\affiliation{\IFPAN}\affiliation{\CSIS} 
\author{L. Scheffler}\affiliation{\Wurtzburg}\affiliation{\ITI} 
\author{C. Gould}\affiliation{\Wurtzburg}\affiliation{\ITI} 
\author{J. Kleinlein}\affiliation{\Wurtzburg}\affiliation{\ITI} 
\author{M. Sawicki}\affiliation{\IFPAN} 
\author{L.W. Molenkamp}\affiliation{\Wurtzburg}\affiliation{\ITI} 
\author{P. Bogus{\l}awski}\email{bogus@ifpan.edu.pl}\affiliation{\IFPAN} 

\date{\today}

\begin{abstract}
High-resolution transmission electron microscopy and superconducting 
quantum interference device magnetometry shows that epitaxial CuMnSb 
films exhibit a coexistence of two magnetic phases, coherently 
intertwined 
in nanometric scales. The dominant $\alpha$~phase is half-Heusler cubic 
antiferromagnet with the N\'{e}el temperature of 62~K, the 
equilibrium structure of bulk CuMnSb. The secondary phase is its 
ferromagnetic tetragonal $\beta$ polymorph with the Curie temperature 
of about 100~K. First principles calculations provide a consistent 
interpretation of experiment, since (i) total energy of $\beta$--CuMnSb 
is 
higher than that of $\alpha$--CuMnSb only by 0.12~eV per formula unit, 
which allows for epitaxial stabilization of this phase, (ii) the metallic 
character of $\beta$--CuMnSb favors the Ruderman-Kittel-Kasuya-Yoshida 
ferromagnetic coupling, and (iii) the calculated effective Curie-Weiss 
magnetic moment of Mn ions in both phases is about 
$5.5~\mu_\mathrm{B}$, favorably close to the measured value. Calculated 
properties of all point native defects indicate that the most likely to 
occur 
are $\mathrm{Mn}_\mathrm{Cu}$ antisites. They affect magnetic properties 
of epilayers, but they cannot induce the ferromagnetic order in CuMnSb. 
Combined, the findings highlight a practical route towards fabrication of 
functional materials in which coexisting polymorphs provide 
complementing functionalities in one host. 

\end{abstract}


\maketitle

\section{Introduction}
\label{sec:intro}

One of the most challenging and long-standing problems in fundamental
magnetism is a competition between ferromagnetic and
antiferromagnetic phases. Their interplay at the interface results
in a well known effect of the exchange
bias,~\cite{Meiklejohn:1957_PR,Nogues:1999_JMMM}
which fuels now a rapid
development of spintronics~\cite{Xiong:2022_FR} and unconventional
computing.~\cite{Borders:2019_N} 
The material class of Heusler alloys was previously used to
study the origin of the transition between magnetic phases because it
offers a wide spectrum of functionalities.~\cite{Wollmann:2017_ARMR} 
Indeed, Heusler alloys exhibit ferromagnetic (FM),  antiferromagnetic
(AFM), and canted ferromagnetic  order.  This indicates that different
types of magnetic coupling are competing in this family. Moreover, some
of its members display structural polymorphism, which allows studying
relationships between the crystalline phase, the magnetic phase, and the
corresponding electronic structure.

Heusler alloys incorporate full-Heusler (X$_2$YZ) and half-Heusler (XYZ)
variants, where X and Y stand for transition metals, whereas Z denotes
anions from the main group. In this class, qualitative changes in
material characteristics can be achieved by chemical substitution on
either the transition metal cation or on the anion sublattice. Typically,
the change of the cation does not change the crystal structure, but it 
can 
induce a crossover between the AFM and the FM magnetic phases. A rarely
met complete solubility with only marginally affected crystallinity of
the otherwise chemically homogenous systems allowed to study the FM-AFM
phase competition in detail. The prominent examples are quaternary solid
solutions such as
Ru$_2$Mn$_{1-x}$Fe$_x$Sn~\cite{Douglas:2016_PRB,Decolvenaere:2019_PRM,
McCalla:2021_PRM} 
Heuslers, and
Co$_{1-x}$Ni$_{x}$MnSb,~\cite{Cong:2010_APL,Yuan:2016_PRB} 
Cu$_{1-x}$Ni$_{x}$MnSb,~\cite{Halder:2011_PRB,Ren_Ni_A,Ren_Ni_B}
Co$_{1-x}$Cu$_{x}$MnSb,~\cite{Duong} and 
Cu$_{1-x}$Pd$_{x}$MnSb~\cite{endo2} half-Heuslers.
In the latter case, the crossover between AFM to FM phases is related to
a change in the electronic structure from semimetallic to 
half-metallic.~\cite{Galanakis, Sasioglu} 

Cu--based CuMnZ compounds are antiferromagnets. This feature attracts
attention given the recent progress achieved in the AFM 
spintronics.~\cite{Jungwirth2018_NP} 
Of particular interest is CuMnAs, with a high
N\'{e}el temperature $T_\mathrm{N}=480$~K.~\cite{wadley} In this case,
features essential for applications, such as anisotropic
magnetoresistance,~\cite{Wang:2020_PRB,poole} current-induced electrical
switching of the N\'{e}el vector~\cite{wadley2016} and of the magnetic
domains,~\cite{grzybowski} have been demonstrated.

The AFM order of CuMnZ is independent of the actual crystalline
structure. The equilibrium structure of bulk CuMnP and CuMnAs is
orthorhombic, while that of CuMnSb is half-Heusler cubic, referred to
below as the $\alpha$~phase. On the other hand, epitaxial growth can
stabilizes metastable phases. This is the case of epitaxial layers of
CuMnAs, grown on both GaP~\cite{wadley, reimers, Wang:2020_PRB, poole,
linn}  and GaAs~\cite{grzybowski} substrates, which crystalize in the
tetragonal structure, referred to below as the $\beta$~phase. Theoretical
investigations of the crystalline properties of CuMnZ series show that
the total energy difference between the cubic and orthorhombic phase is
about 1~eV per f.u. (formula unit) for CuMnP, and about 0.5~eV per f.u. 
for
CuMnAs.~\cite{maca_JMMM} This suggests that the orthorhombic phase of
CuMnSb, the last member of the CuMnZ series, may not be stable, and
indeed the stable structure is the $\alpha$~phase. However, as we show
here, epitaxial stabilization of CuMnSb in the $\beta$~phase is in
principle possible, because the calculated energy difference between
$\alpha$--CuMnSb and $\beta$--CuMnSb is small, about 0.12~eV per f.u.
Moreover, the $\beta$--CuMnSb polymorph becomes stable at pressures above
7~GPa.~\cite{Malavi} 

Concerning the magnetic properties, the N\'{e}el temperature of both
orthorhombic CuMnAs and $\beta$--CuMnAs is well above the room
temperature,~\cite{wadley} whereas that of $\alpha$--CuMnSb is lower,
about 60~K.~\cite{Endo1968_PSJap,LukasaAPL} 
Theory agrees with experiment, since according to
Ref.~\onlinecite{maca_condmat}, in the orthorhombic
CuMnP and CuMnAs, the AFM order is more stable than the FM by about
250~meV/Mn.
This energy difference is smaller in the cubic phase of CuMnZ compounds,
for which the AFM order is lower in energy than
FM one by about 50~meV per f.u.~\cite{Pickett, maca_condmat} 
Finally, the AFM order of $\alpha$--CuMnSb is stable under applied
magnetic field, as $T_\mathrm{N}$ does not change up to
50~Tesla.~\cite{doerr}

Turning to the electronic structure of the CuMnP-CuMnAs-CuMnSb series we
observe that the character of the energy band gap
depends on the anion.
Similar to the case of e.g. zinc blende semiconductors, the band gap
decreases with the increasing atomic number of the
anion.~\cite{maca_condmat} 
Indeed, CuMnP is a semiconductor, CuMnAs  has a practically vanishing
band gap, and CuMnSb is a semimetal.~\cite{comment}

Here we experimentally confirm a puzzling coexistence of AFM and FM
phases in epitaxial stoichiometric CuMnSb films, observed by us
previously,~\cite{PRM} and explain the underlying mechanism responsible
for this effect. A fine analysis of transmission electron microscopy 
(TEM)
images, Sec.~\ref{sec:tem},  points to the formation of tetragonal 
$\beta$--CuMnSb inclusions embedded coherently within the cubic 
$\alpha$--CuMnSb host. 
The tetragonal structure of these inclusions is the same as
that of the tetragonal $\beta$--CuMnAs. Magnetic properties of our films,
Sec.~\ref{sec:magnet}, demonstrate coexistence of two magnetic phases:
apart from the dominant AFM one, expected for CuMnSb, the
measurements reveal the presence of a FM contribution. This is an
unexpected feature within the CuMnZ series, exhibiting the AFM order.

In Sec.~\ref{sec:theory}, we employ calculations based on the density
functional theory to assess properties of CuMnSb films. In agreement with
the experiment, $\beta$--CuMnSb is weakly metastable, but its magnetic
ground state is FM. Band structures of $\alpha$ and $\beta$ polymorphs
are close, but changes in the density of states at the Fermi level
account for the change of the dominant mechanism
of the magnetic coupling from AFM superexchange to FM Ruderman-Kittel-
Kasuya-Yoshida (RKKY).
Finally, in Sec.~\ref{sec:eform} native point defects in CuMnSb are
examined to assess their possible influence on the magnetic
properties.~\cite{Maca94} Our results indicate that the dominant native
defects in $\alpha$--CuMnSb are Mn antisites, and their presence in the
films can possibly account for small differences between the measured and
the calculated magnetic characteristics, but they do not stabilize the FM
order of $\alpha$--CuMnSb.


\section{Experimental Results}
\label{sec:exp}

\begin{figure*}[ht!]
\includegraphics[width=0.95\textwidth]{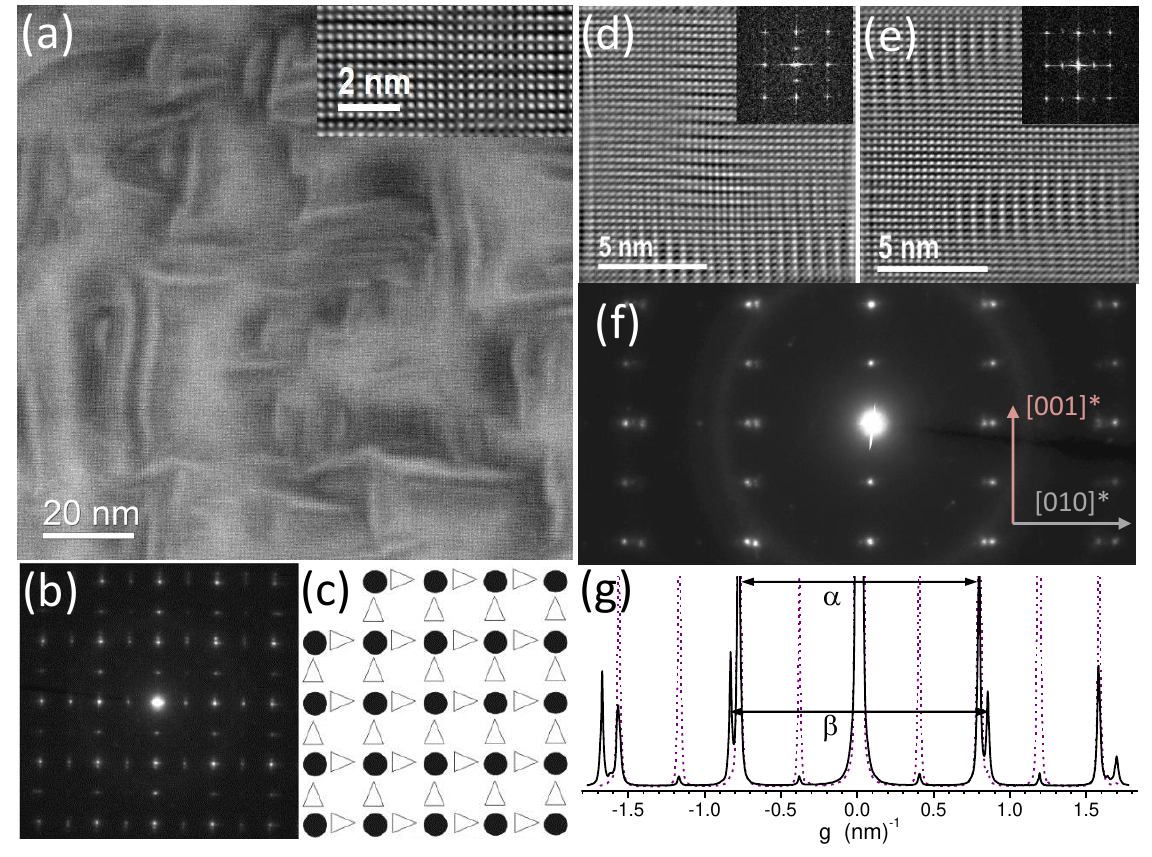}
\caption{(a) High-angle annular dark-field scanning transmission electron
microscopy image of a CuMnSb layer in the [100] zone axis. The inset in
the top-right corner brings up a part of the image in atomic resolution,
where bright dots represent columns of Mn and Sb atoms. (b) Electron
diffraction pattern of the layer. (c) Schematics of the positions of
Bragg's spots from (b). Big bullets represent the main reflections from
the cubic CuMnSb structure, whereas the open triangles mark the positions
of the weak extra reflections. The orientation of the triangles follows
from the analysis of the data in panels (d-g). (d-e) Blown up two
regions from (a), in which either vertical or horizontal strips dominate.
The corresponding Fourier transforms are showed in the top right corners
of both panels. (f) Selected area electron diffraction  pattern taken at
the regions dominated by the vertically oriented strips. (g) Diffraction
intensity profiles taken along the horizontal [010]* and the vertical
[001]* lines passing through the center the diffraction pattern. The
solid line corresponds to the horizontal [010]* direction and the dashed
one to the vertical [001]* one in panel (f).
Stars denote directions in the reciprocal space.
The  arrows $\alpha$ and
$\beta$ indicate the length of $\alpha$--2g(002) and $\beta$--2g(002)
diffraction vectors, respectively. }
\label{fig:TEM1}
\end{figure*}

\subsection{Experimental Methods}
\label{sec:Exp-Sec}
{\bf Growth conditions. }

CuMnSb layers about 200 nm thick are grown by molecular beam
epitaxy. 
Separate growth chambers connected by an ultra-high vacuum
transfer system are used for the growth of the individual layers.
Low tellurium-doped epi-ready GaSb (001) wafers are used 
as substrates. 
Prior to the growth, the natural oxide layer is desorbed in an Sb
atmosphere.
Then, 150~nm thick GaSb buffer layers are grown on the substrates to
ensure a 
high-quality interface for the growth of CuMnSb. The GaSb buffer layers
are grown at a substrate temperature of $530^{\circ}\mathrm{C}$ and a
beam equivalent pressure of $4.0\times 10^{-6}$~mbar and 
$5.3\times 10^{-7}$~mbar for Sb and Ga, respectively. 
Sb supply is facilitated by a single-filament effusion cell, while Ga is 
provided by a double-filament effusion cell.

A substrate temperature of $250^{\circ}\mathrm{C}$ is used for the growth
of CuMnSb films. The corresponding beam equivalent pressures are as
follows: $\mathrm{BEP}_\mathrm{Cu} = 5.80 \times 10^{-9}$~mbar,
$\mathrm{BEP}_\mathrm{Mn} = 9.03\times 10^{-9}$~mbar, and
$\mathrm{BEP}_\mathrm{Sb} = 4.23\times 10^{-8}$~mbar. Cu is supplied by a
double filament effusion cell, while Mn and Sb are supplied by single
filament effusion cells.
Following the growth of CuMnSb, a 2.5~nm thick  layer of Al$_2$O$_3$ is
deposited on the samples through a sequential process of aluminum DC
magnetron sputtering and oxidation.
Please, refer to Ref.~\onlinecite{LukasaAPL} for a comprehensive analysis 
of the growth process and physical
properties of the CuMnSb layers produced using the methodology outlined
above. 

{\bf Transmission Electron Microscopy. }
Specimens for the transmission electron microscopy (TEM) investigations
are
prepared by the focused ion beam method in the form of lamellas cut along
the [100] and [110] directions, i.e., perpendicularly to the surface
(001) plane.
Titan Cubed 80-300 electron transmission microscope operating with
accelerating voltage 300~kV and equipped with energy-dispersive X-ray
spectrometer (EDXS) is used for the study. Most of the investigations are
done
on Cu grids, but for EDXS elemental analysis a Mo grid is used to avoid
interference of Cu fluorescence signal from the grid. This analysis
yields
percentage atomic concentration at 37(3)\,:\,32(5)\,:\,31(7) for Cu, Mn,
and
Sb, respectively, which, within the experimental errors (given in the
parentheses), correspond to the expected stoichiometric ratio of
33\,:\,33\,:\,33.

{\bf SQUID Magnetometry. }
Magnetic characterization is performed in a commercial superconducting
quantum interference device (SQUID) magnetometer MPMS XL7. The magnetic
moment of antiferromagnetic layers is generally very weak and by far
dominated by the magnetic response of the bulky semiconductor substrates.
Therefore, to counter act the typical shortcomings of commercial
magnetometers built 
around superconducting magnets~\cite{Sawicki_2011} and to minimize
subtraction errors during data reduction we actively employ the in situ
compensation.~\cite{Gas_2022} It allows us to reduce the coupling of the
signal of the substrates to about 10\% of their original strength.
The actual effectiveness of the compensation depends on the mass of the
sample and 
its orientation with respect to the SQUID pick-up
coils.~\cite{Sawicki_2011, Stamenov} 
We also strongly underline the importance of
a thorough mechanical removal of the metallic MBE glue from the backside
of the samples for any magnetic studies.
Its strongly nonlinear magnetic
contribution can be of the same magnitude as that of the layer of
interest.~\cite{Gas_2021} 
To accurately establish the magnitude of magnetic moment 
specific to CuMnSb we measure a reference sample grown without the
CuMnSb layer~\cite{LukasaAPL} using the same sample holder and following
exactly the same experimental
sequence as that executed for the investigated samples.

\subsection{Structural characterization }
\label{sec:growth}


\label{sec:tem}
An exemplary atomic resolution high-angle annular dark-field scanning
transmission electron microscopy (HAADF/STEM) 
image obtained for the [100]
zone axis (the direction of the projection) is in
{Fig.~\ref{fig:TEM1}}~(a).
It confirms a high quality cubic constitution of the material, as it is
underlined in
the inset. However, at the contrast chosen here, the image in this field
of view
reveals the presence of stripe-like features, which are the main subject
of this
analysis. In this image, the apparent lengths and widths of the strips
are about 40~nm and about 4~nm, respectively, running predominantly 
either vertically 
or horizontally in this particular projection. 
On other images, the strips exhibit a
relatively wide distribution of lengths in the 10-100 nm window. Since 
similarly
distributed shadowy stripes are observed also in the [110] zone axis, we
conclude that they form along all three principal crystallographic
directions
without any particular preferences. The expected F$\overline{4}3$m cubic
structure of $\alpha$--CuMnSb is clearly confirmed by the fourfold
symmetry of
the dominant (bright) spots seen on electron diffraction pattern
presented in
Fig.~\ref{fig:TEM1}~(b).

Importantly, the diffraction pattern in Fig.~\ref{fig:TEM1}~(b)
contains also a
second set of much fainter reflections, situated halfway between two
adjacent
reflections of the main pattern. This indicates the presence of a second
crystallographic $\beta$~phase, which periodicity in the corresponding
direction
is doubled relative to that of $\alpha$--CuMnSb, but otherwise coherent
with this
host structure. We bring all the Bragg's spots up in
Fig.~\ref{fig:TEM1}~(c), in
which the bullets represent the main reflections from $\alpha$--CuMnSb,
whereas the open triangles mark the positions of the weak ones, which are
forbidden for this structure.

The presence of $\beta$--CuMnSb is further substantiated by the
inspection of
the two close-ups from Fig.~\ref{fig:TEM1}~(a), shown in
Fig.~\ref{fig:TEM1}~(d)
and (e). At this magnification they reveal that, on top of the otherwise
cubic
arrangement of atomic columns, the strips' brightness alternates every
second
$\{002\}$ plane along the direction perpendicular to strip's long axis.
The
modulation is vertical in Fig.~\ref{fig:TEM1}~(d), whereas it goes
horizontally in
Fig.~\ref{fig:TEM1}~(e). The top right corners of these figures contain
the
corresponding Fourier transform of the parent image, and, similarly to
Fig.~\ref{fig:TEM1}~(b), both patterns are dominated by the main
reflections of
$\alpha$--CuMnSb. The additional spots are embedded either along vertical
[Fig.~\ref{fig:TEM1}~(d)] or horizontal [Fig.~\ref{fig:TEM1}~(e)]
lines, i.e., the
presence of vertical and horizontal orientations is mutually exclusive.
This feature is reflected in Fig.~\ref{fig:TEM1}~(c), 
where the additional spots are
marked by differently oriented triangles. The triangles with apexes
directed vertically correspond to the vertical orientation of 
the brightness modulation in
Fig.~\ref{fig:TEM1}~(d), whereas the horizontal direction of apexes
corresponds to the horizontal modulation.

\begin{figure}[t]
\includegraphics[width=0.5\textwidth]{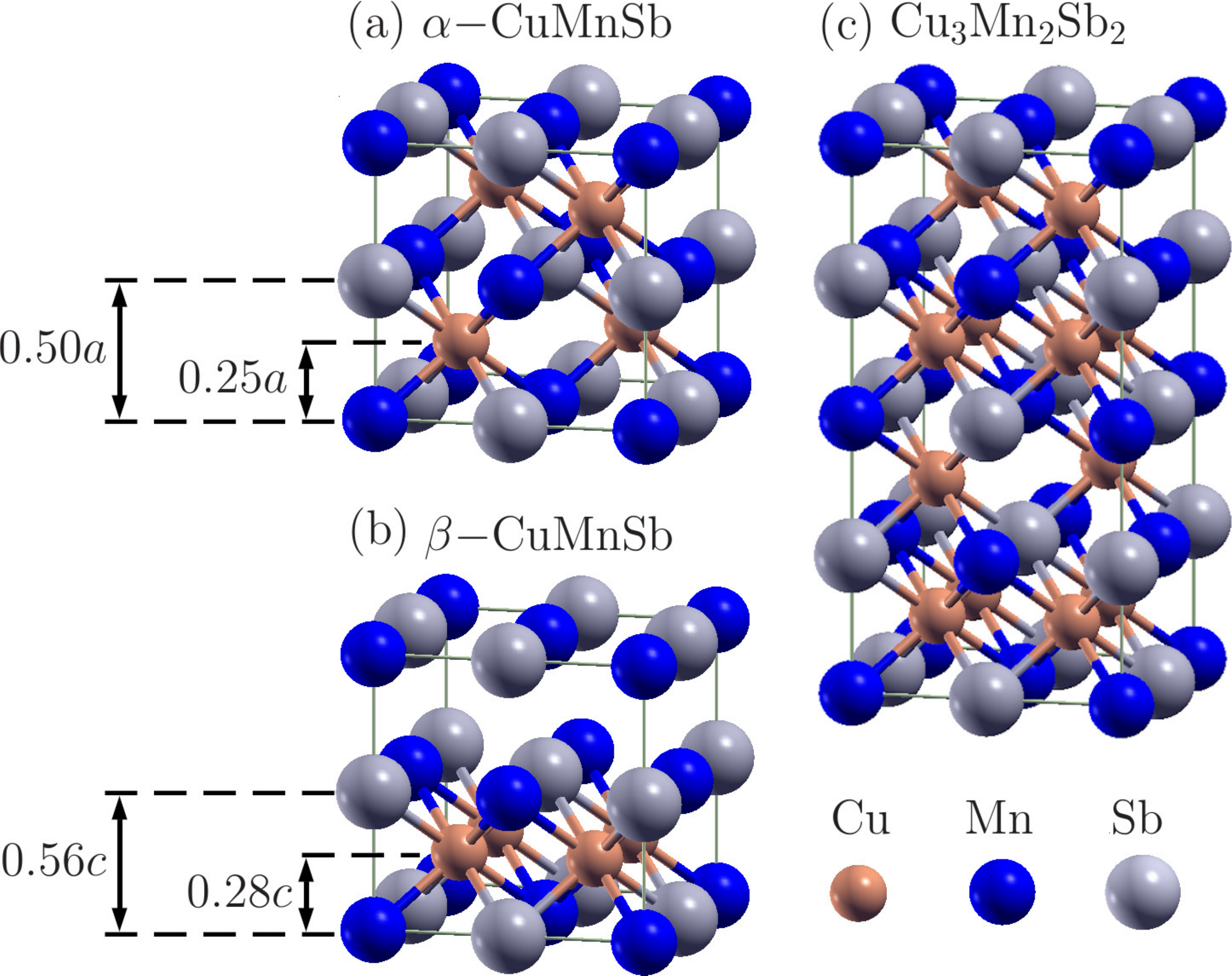}
\caption{Crystal structures of (a) $\alpha$--CuMnSb with the cubic
lattice
constant $a$, (b) tetragonal $\beta$--CuMnSb with the lattice constants
$a$ in the
($x$, $y$) plane and $c$ in the [001] direction, and (c)
Cu$_3$Mn$_2$Sb$_2$.}
\label{fig:str_PD}
\end{figure}

Based on the data shown above we propose that the second phase of
CuMnSb, present in our films in the form of strips, is a tetragonal
structure,
which also is the structure of epitaxial CuMnAs,~\cite{wadley, reimers,
Wang:2020_PRB, poole, linn, grzybowski} and of CuMnSb at high
pressures.~\cite{Malavi} This $\beta$--CuMnSb polymorph is shown in the
panel (b) of {Fig.~\ref{fig:str_PD}}. The difference between
$\alpha$ and
$\beta$~phases consists in the location of Cu ions: in the $\alpha$ phase
every
(001) plane between two consecutive MnSb planes is half-occupied by Cu,
whereas in the $\beta$~phase Cu ions completely fill up every second
(001)
plane, and the overall stoichiometry of the material is preserved.

Details regarding $\beta$--CuMnSb can be inferred from selected area
electron diffraction (SAED) patterns taken at regions with different
orientations of the strips. Diffraction pattern of an area dominated by
the
vertically oriented strips is shown in more detail in
Fig.~\ref{fig:TEM1}~(f). In
agreement with the Fourier transforms, SAED shows the occurrence of
specific reflections corresponding to this particular orientation. 
The reflections common to both the cubic $\alpha$ and the tetragonal
$\beta$ polymorphs are split along the [010]* direction, i.e., orthogonal
to
the strip's axis, whereas the weak spots of the $\beta$~phase are not
split and
are commensurate with the cubic phase. (A star denotes a direction in the
reciprocal space.)

We quantify the effect analyzing intensity profiles taken along lines
passing
through the center of diffraction. The profiles are superimposed, and
presented in Fig.~\ref{fig:TEM1}~(g). The profile along the [001]*
direction
reflects the periodicity of $\alpha$--CuMnSb, while that along [010]* is
additionally split. From the Figure it follows that in our specimens the
$c$
lattice parameter of the $\beta$--CuMnSb strips is equal to that of the
host
$\alpha$--CuMnSb, 6.2(1)~{\AA}, whereas the $a$ and $b$ parameters of the
$\beta$~phase, 5.8(1)~{\AA}, are smaller by about 7\%. Analogous features
are observed for the [010]-oriented strips.

\begin{figure}[t]
\includegraphics[width=0.5\textwidth]{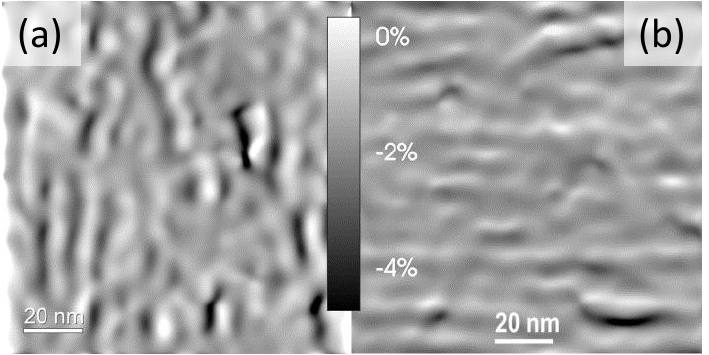}
\caption{
Strain maps of image shown in Fig.~\ref{fig:TEM1}~(a).
(a) The horizontal component of strain $\epsilon_{xx}$, and
(b) the vertical one, $\epsilon_{zz}$.
Geometrical phase analysis method has been applied.~\cite{GPA:2018} 
}
\label{fig:TEM2}
\end{figure}

The existence of such a significant strain is confirmed by the
calculation of
strain maps. We apply the geometrical phase analysis
method~\cite{GPA:2018}
for the main image presented in Fig.~\ref{fig:TEM1}~(a),
and the results are presented in {Fig.~\ref{fig:TEM2}}~(a) 
and (b) for the
horizontal, $\epsilon_{xx}$, and the vertical, $\epsilon _{zz}$,
components of strain, respectively. It is seen that stripes' 
strain is negative (dark shade)
perpendicular to strips and almost zero along the strips. For example, on
the horizontal strain map [Fig.~\ref{fig:TEM2}~(a)] only vertical
strips are visible because they are compressed horizontally, 
whereas the horizontal strips are invisible because they are 
not deformed in the horizontal
direction.

The calculated properties of $\beta$--CuMnSb, such as its lattice
parameters, stability, and magnetic properties, are discussed in detail
in Sec.~\ref{sec:beta}. 
Anticipating, we mention that they are consistent with
experiment. We have also considered a second possible structure which is
(almost) compatible with the TEM data, Cu$_3$Mn$_2$Sb$_2$, depicted in
Fig.~\ref{fig:str_PD}~(c). However, this compound is 
higher in energy than the $\beta$~phase, and was dropped 
from further considerations.


\subsection{Magnetic properties}
\label{sec:magnet}

\begin{figure}[t]
\includegraphics[width=0.5\textwidth]{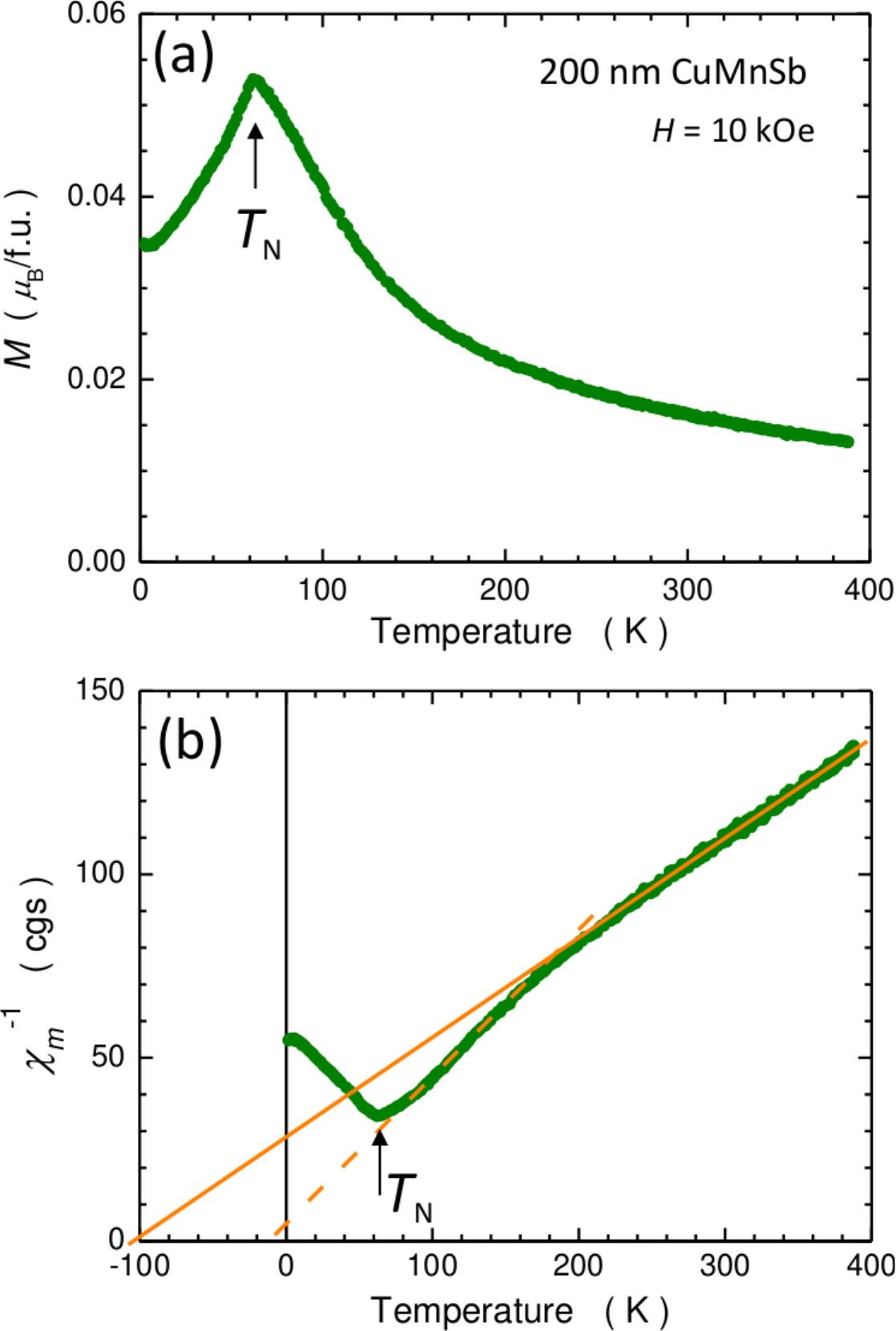}
\caption{
Results of temperature dependent magnetic studies of 200~nm thick CuMnSb 
layer 
(green bullets). 
(a) Magnetization $M$ established in a bias field of $H=10$~kOe, and 
(b) the corresponding inverse of the molar magnetic susceptibility 
$\chi_m^{-1}$. The solid and dashed 
orange  lines indicate the Curie-Weiss behavior 
of $\chi_m^{-1}(T)$ for $T > 200$~K
and for $T_\mathrm{N} < T < 200$~K, respectively. 
The arrows indicate the position of the N\'{e}el temperature, 
$T_\mathrm{N} = 62$~K, 
on both panels.}
\label{fig:M1}
\end{figure}

The temperature $T$ dependence of magnetization, $M(T)$, of the 200 nm
thick layer of CuMnSb, is depicted in {Fig.~\ref{fig:M1}} (a).
The clear kink on $M(T)$ at $T_\mathrm{N} = 62$~K marks the position of
the paramagnetic to antiferromagnetic N\'{e}el transition in the layer.
This value corresponds precisely to the values of $T_\mathrm{N}$
established previously for CuMnSb/GaSb layers of the thickness 
$t \geq 200$~nm, what, indirectly, 
indicates stoichiometric material composition of this
layer.~\cite{LukasaAPL} 

More specific information about the magnetic state of that sample is 
obtained from the examination of the temperature dependence of the 
inverse magnetic susceptibility, $\chi^{-1}(T)$, shown in 
Fig.~\ref{fig:M1} 
(b). We take here $\chi(T) = M(T)/H$, where $H=10$~kOe is the external 
magnetic field applied during the measurements. $\chi^{-1}(T)$ can be 
approximated by two straight lines. The abscissa of the first one, which 
approximates $\chi^{-1}(T)$ above 200~K (the solid orange line in 
Fig.~\ref{fig:M1}), yields exactly the same magnitude of the 
Curie-Weiss 
temperature $T_\mathrm{CW} =-100(5)$~K as that established 
previously for a thicker 510~nm layer, for which $\chi^{-1}(T)$ formed a 
single straight line above $T_\mathrm{N}$ at the same experimental 
conditions.~\cite{LukasaAPL} Also the slope of this line yields the value 
of 
the effective magnetic moment $m_{\mathrm{eff}} = 5.4(1) 
\mu_\mathrm{B}$ per f.u., which is very close to that found previously, 
$m_{\mathrm{eff}} = 5.6 \mu_\mathrm{B}$ per f.u.~\cite{LukasaAPL} This 
correspondence indicates that the high temperature part of $\chi^{-1}(T)$ 
is 
determined predominantly by AFM excitations in the paramagnetic matrix of 
CuMnSb. 

The abscissa of the second straight line, which approximates the 
experimental data between $T_\mathrm{N}$ and about 200~K (marked as 
the dashed orange line in Fig.~\ref{fig:M1}), yields a more positive 
value 
of the Curie-Weiss temperature, $T_{CW}' = -10(10)$~K. This clear 
positive shift of $T_{CW}$ indicates the existence of a 
ferromagnetic 
contribution to the overall antiferromagnetic phase of the material, and 
that 
these FM excitations gain in importance below about 200~K. Interestingly, 
a 
somewhat stronger effect, characterized by a change of sign of 
$T_{CW}$ to $T_\mathrm{CW}' = +60(10)$~K, was noted in 
40~nm CuMnSb layer grown on InAs.~\cite{PRM} In accordance with the 
findings of structural characterization we propose that the by far 
stronger 
AFM component originates from the dominant $\alpha$~phase, whereas the 
FM one is brought about by $\beta$--CuMnSb polymorph. 

Turning now to the magnetic characteristics established here for 
$\alpha$--CuMnSb we note that they are close to those reported 
previously, as 
shown in Tab.~\ref{tab:TNeel}. The published data exhibit a 
certain distribution, which may indicate that other factors, such as  a 
weak 
crystalline disorder, may be at work. In particular, either additional Mn 
interstitial ions or Cu$_\mathrm{Mn}$-Mn$_\mathrm{Cu}$ antisite pairs are 
likely 
to form.~\cite{Maca94} The presence of such defects was suggested to 
stabilize the experimentally observed AFM \{111\}-oriented phase of 
$\alpha$--CuMnSb.~\cite{Maca94} Finally, we do not observe a canted 
AFM order at low temperatures~\cite{Regnat} in any of our samples.

\vspace{0.5cm}
\begin{table}
\caption{Experimental N\'{e}el temperature $T_\mathrm{N}$, effective
Curie-Weiss magnetic moment of Mn ions $m_\mathrm{eff}$(Mn),
and Curie-Weiss temperature $T_\mathrm{CW}$ of $\alpha$--CuMnSb.
Measured orientation of the AFM axis is also given (n.e.~= not
established). Refs.~\onlinecite{Forster}
and~\onlinecite{Regnat} report the saturation Mn moment. }
\label{tab:TNeel}
\begin{ruledtabular}
\begin{tabular}{lllcc}
$T_\mathrm{N}$ (K) & $m_\mathrm{eff}$(Mn) ($\mu_\mathrm{B}$) & 
$T_\mathrm{CW}$ (K)
& AFM order & Ref. \\
\hline
-   & 3.9(1) & - & [111] & \onlinecite{Forster}\\
55  & 3.95 & -160(8) & [111] & \onlinecite{Regnat}\\
55  & 5.4 & -160 & [111] & \onlinecite{Endo}\\
62 	& 5.2 & -120 &   n.e.   & \onlinecite{Helmholdt}\\
50  & 6.3 & -250 &   n.e.   & \onlinecite{Boeuf}\\
50  & - & - & n.e. & \onlinecite{maca_JMMM}\\
62  & 5.9 &  -65 & n.e. & \onlinecite{PRM}\\
62  & 5.6 & -100 & n.e. & \onlinecite{LukasaAPL}\\
62  & 5.5 & -100 & n.e. & this work\\
\end{tabular}
\end{ruledtabular}
\end{table}

\section{Theory}
\label{sec:theory}

\subsection{Theoretical Methods}
Calculations are performed within the density functional
theory~\cite{Hohenberg,KohnSham} in the generalized gradient
approximation of the exchange-correlation potential proposed
by Perdew, Burke and Ernzerhof.~\cite{PBE} To improve description
of $3d$ electrons, the Hubbard-type $+U$ correction on Mn is
added.~\cite{Anisimov1991, Anisimov1993, Cococcioni} The
parameter $U(\mathrm{Mn})=1$~eV reproduces the known formation
energy of the intermetallic CuMn alloy and gives a reasonable value 
of the Mn cohesive
energy. We use the pseudopotential method implemented in the
{\sc Quantum ESPRESSO} code,~\cite{QE} with the valence atomic
configuration $4s^{1.5} p^0 3d^{9.5}$ for Cu, \\
$3s^2 p^6 4s^2 p^0 3d^5$ for Mn and $5s^2 p^3$ for Sb ions. 
The plane-waves kinetic
energy cutoffs of 50~Ry for wave functions and 250~Ry for charge
density are employed. Finally, geometry relaxations are performed
with a 0.05~GPa convergence criterion for pressure. In defected
crystals ionic positions are optimized until the forces acting on
ions become smaller than 0.02~eV/\AA.

The properties of defected $\alpha$--CuMnSb are examined using
cubic $2a\times 2a\times 2a$ supercells with 96 atoms (i.e., 32 f.u.),
while magnetic order of ideal crystals are checked using the smallest
possible supercells. Here $a$ is the equilibrium lattice parameter.
The $k$-space summations are performed with a $6\times 6\times
6$ $k$-point grid for the largest supercell, and correspondingly
denser grids are used for smaller cells.

Magnetic interactions and magnetic order depend on several factors, such
as the exchange spin splitting of the $d$(TM) shells, charge states of TM
ions, concentration of free carriers and their spin polarization, and the
density of states (DOS) at the Fermi energy $E_\mathrm{F}$. These factors
are interrelated, and are calculated self-consistently within ab initio
approach.

Considering first the localized magnetic moments we note that spin
polarization of Co, Ni, and Cu ions in XMnZ compounds practically 
vanishes,
while that of the $d$(Mn) shell is substantial.~\cite{Pickett, Galanakis,
Sasioglu} The robustness of the Mn magnetic moment results from the
large, 3 -- 5 eV, spin splitting of the $3d$(Mn) states. In fact, in XMnZ 
the
$d$(Mn) spin up channel is occupied, while most of the spin down $d$(Mn)
states lay above the Fermi level. Here, one can observe that spin
polarization of the $d$(TM) electrons in free atoms depends on the
difference in the number of spin up and spin down electrons, which is the
highest in the case of Mn. Consequently, the Mn spin polarization 
persists in
XMnZ. On the other hand, spin splitting of $d$ electrons of Co and 
Ni atoms is smaller, and thus 
it vanishes in XMnSb hosts, see the analysis for TM dopants in
ZnO.~\cite{ciechan} 

In CuMnSb, the magnetic sublattice consists of Mn ions, which are second
neighbors distant by 4.3~\AA. Therefore, the direct exchange coupling
between two Mn ions, given by overlaps of their $d$(Mn) orbitals, is
negligibly small. The remaining indirect exchange coupling is the sum of
two contributions, and the exchange constant 
$J_{indirect} = J_{sr} +J_{RKKY}$.~\cite{Falicov, Galanakis, Sasioglu}  
The first term $J_{sr}$ has a
short-range AFM character, and it is inversely proportional to the energy
distance between the unoccupied $d$(Mn) states and $E_\mathrm{F}$. The
second coupling channel is of RKKY type mediated by free carriers. This
channel depends on the detailed electronic structure in the vicinity of
$E_\mathrm{F}$, and $J_{RKKY}$ is proportional to DOS$(E_\mathrm{F})$. In
particular, CoMnSb and NiMnSb half-metals are FM, while CuMnP and
CuMnAs insulators are AFM. As we show here, CuMnSb is the border case.

\subsection{Crystal and magnetic properties of $\alpha$--CuMnSb}
\label{sec:alpha}

\begin{figure}[t]
\includegraphics[width=0.5\textwidth]{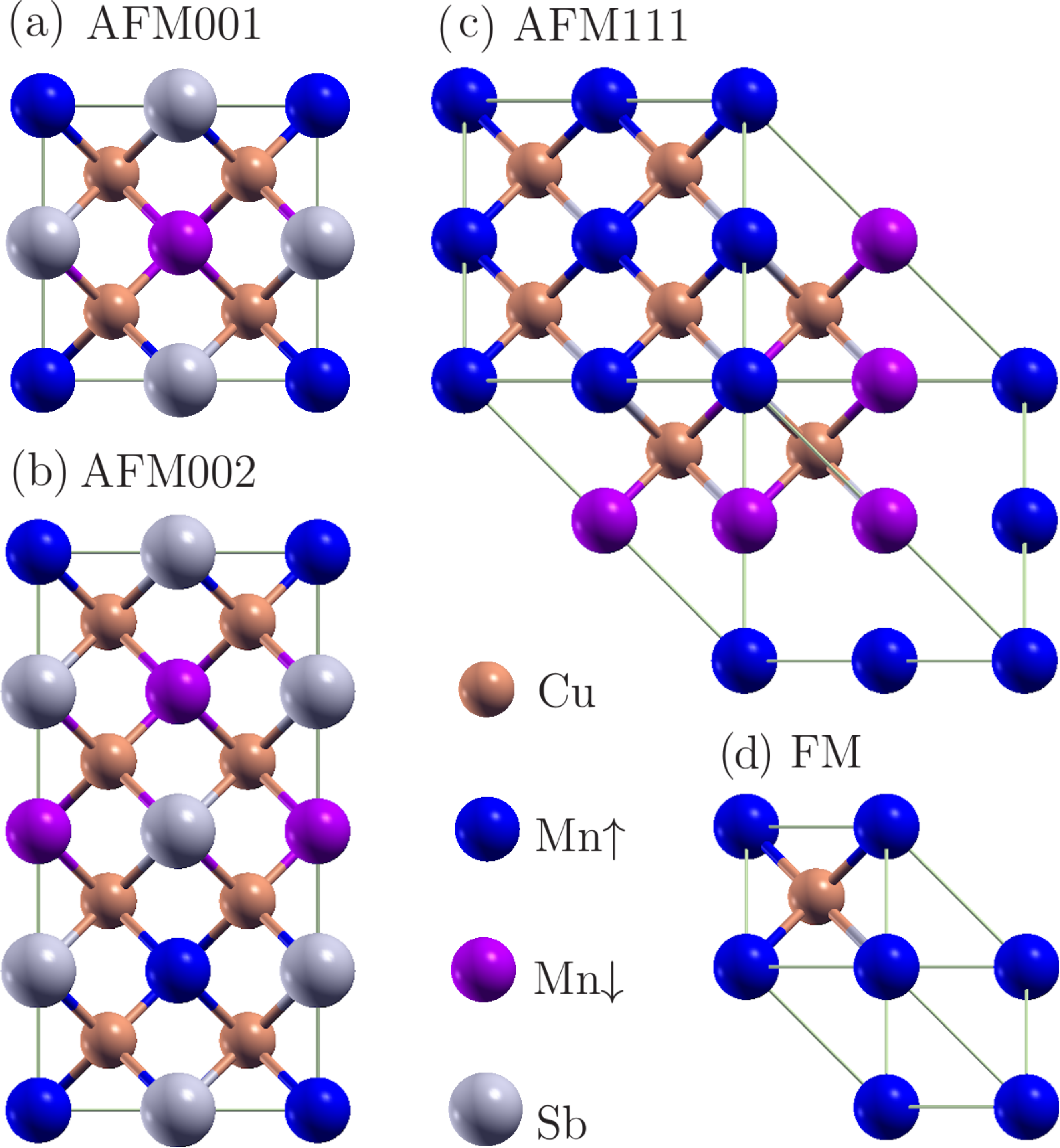}
\caption{Side view of the considered
magnetic cells:
(a) antiferromagnetic with \{001\} planes with the same Mn
spins shown in the $a\times a\times a$ cell (AFM001), (b)
antiferromagnetic
with double \{001\} planes of the same spins on Mn ions in the $a\times
a\times 2a$ cell (AFM002), (c) antiferromagnetic with ferromagnetic
\{111\} planes realized in the $a\sqrt{2}\times a\sqrt{2}\times
a\sqrt{2}$ cell (AFM111),  and (d) ferromagnetic in primitive
rhombohedral cell
$a/\sqrt{2}\times a/\sqrt{2}\times a/\sqrt{2}$ (FM).
Mn atoms with different spin directions are indicated as Mn$\uparrow$
and Mn$\downarrow$.}
\label{fig:mag}
\end{figure}

A rhombohedral primitive cell of $\alpha$--CuMnSb contains 
one formula unit. This structure
consist in four interpenetrating fcc sublattices, one of them being 
empty.
The consecutive (001) MnSb planes are followed by the "half-empty" Cu
planes, in which the planar atomic density is twice lower. The cubic unit
cell is presented in Fig.~\ref{fig:str_PD}~(a). Local 
coordination of
Mn ions can be relevant from the point of view of magnetic interactions.
With this respect we notice that the magnetic coordination of an Mn ion
consists in 12 equidistant Mn atoms at $a/\sqrt{2}$. Moreover, the 
short-range coupling between two Mn nearest neighbors is mediated through
the closest ions, and it occurs through  one Mn-Cu-Mn "bridge" and two
Mn-Sb-Mn "bridges".

We consider four magnetic phases of $\alpha$--CuMnSb. The
corresponding supercells are shown in Fig.~\ref{fig:mag}.
Antiferromagnetic order with parallel Mn spins in the (001) planes,
AFM001, is calculated using the cubic $a\times a\times a$ cell with 
4~f.u.
(12 atoms), and shown in Fig.~\ref{fig:mag}~(a). The AFM order with a
period doubled in the [001] direction with parallel Mn spins in each 
(001)
plane, denoted as AFM002, is shown in Fig.~\ref{fig:mag}~(b). The
corresponding $a\times a\times 2a$ cell contains 8~f.u., and is one of 
the
possible supercells in which this phase can be realized. In the AFM111
phase, the Mn spins are parallel in each (111) plane, but the consecutive
(111) planes are AFM, as shown in Fig.~\ref{fig:mag}~(c), and the
corresponding rhombohedral unit cell 
$a\sqrt{2}\times a\sqrt{2}\times a\sqrt{2}$ contains 8 primitive cells 
with 24 atoms. Finally, the FM phase requires a primitive cell 
$a/\sqrt{2}\times a/\sqrt{2}\times a/\sqrt{2}$ with 1~f.u., 
presented in Fig.~\ref{fig:mag}~(d).

\begin{figure}[t]
\includegraphics[width=0.5\textwidth]{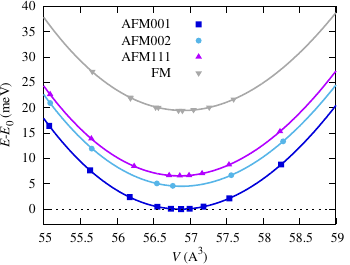}
\caption{Volume dependence of the total energy relative to the ground
state
$E_0=E_{AFM001}(V_0)$ of $\alpha$--CuMnSb in the AFM001, AFM002, AFM111
and FM phases.
Both volume and energy are per formula unit. 
Lines are fitted to the calculated values (symbols). 
}
\label{fig:E-V}
\end{figure}

\begin{table}
\caption{
The calculated lattice parameter $a$, the saturation Mn
magnetic moment, $m_\mathrm{sat}$, and the energy of the given
magnetic order relative
to $\alpha$--CuMnSb in the AFM001 ground state, $\Delta E_{tot}$.
All energies are per one formula unit.
Our measured TEM values are also given.
}
\begin{ruledtabular}
\begin{tabular}{lcccc}
&$a$ (\AA) &$c$ (\AA) & $m_\mathrm{sat} (\mu_\mathrm{B})$ &$\Delta 
E_{tot}$ (meV)
\\
\hline
$\alpha$--CuMnSb\\
AFM001   & 6.10 & - & 4.59 & 0\\
AFM002   & 6.09 & - & 4.61 & 3\\
AFM111   & 6.10 & - & 4.55 & 7\\
FM       & 6.11 & - & 4.70 & 19\\
TEM      & 6.2(1) & - & - & -\\
\hline
$\beta$--CuMnSb\\
AFM   & 5.83 & 6.40	& 4.48 & 113\\
FM   & 5.88 & 6.28	& 4.52 & 102\\
TEM   & 5.8(1) & 6.2(1)	& - & -\\
\end{tabular}
\end{ruledtabular}
\label{tab:mag}
\end{table}

The obtained results are collected in Tab.~\ref{tab:mag} and in
{Fig.~\ref{fig:E-V}}. The ground state structure is AFM001, but 
the
experimentally observed AFM111 is only  7 meV per f.u. higher in energy. 
The
least stable is the FM order, higher in energy than AFM001 by about 20
meV per f.u. 
The equilibrium lattice parameters $a\approx~6.1$~\AA\ are practically
independent of the magnetic order, and close to the experimental
value 6.088 \AA.~\cite{Forster}
Some phases are characterized by a small distortion 
of
the cubic symmetry caused by different bond lengths between
ferromagnetically and antiferromagnetically oriented Mn ions. Differences
in the lattice parameters between various magnetic phases are below 0.01
\AA, and are not reported in the Table. Similar results for the AFM001
order were obtained in Ref.~\onlinecite{Maca94}, while in 
Refs~\onlinecite{Pickett,
Galanakis} the AFM order is more stable than FM by 50 and 90~meV per Mn,
respectively.

The last property reported in Tab.~\ref{tab:mag} is the saturation
magnetic moment of Mn, which also is similar in all phases, and equal to
about $4.6 \mu_\mathrm{B}$. This value corresponds to the Curie-Weiss
moment of $5.5(1) \mu_\mathrm{B}$, and compares favorably with the
experimental values given in Tab.~\ref{tab:TNeel}.

The obtained results allow estimating the relative roles of the short- 
and
long-range contributions to the magnetic coupling. To this end, we assume
the hamiltonian in the form $H_{ex}=-J/2\sum_{i,j}\vec s_i \vec s_j$, 
where
the short range interaction is limited to the Mn NNs neighbors, and the
long-range term is neglected. The spin value, $s_i \approx 2.3$, is one 
half
of the calculated magnetic moment of Mn.

The exchange constant $J$ is positive (negative) for the FM (AFM) 
coupling,
and is obtained by comparing energies of various magnetic orders. In the
AFM001 phase, each Mn ion has 4 ferromagnetically oriented Mn NNs in
the (001) plane and 8 antiferromagnetically oriented Mn NNs in the two
adjacent planes. For the remaining magnetic phases, the energies
calculated relative to the ground state $E_0\equiv E_\mathrm{AFM001}$
depend on the magnetic order as shown in Tab.~\ref{tab:mag}.
These results give the coupling constant in the range 
$-0.6 \ge J_{sr} \ge -0.2$~meV. 
This spread is quite large and cannot be explained by
(negligible) changes in atomic distances in cells with different magnetic
ordering. Therefore, we conclude that the Heisenberg nearest neighbor
model does not describe magnetic properties of $bulk$ phases. Indeed,
such a model is not appropriate for metallic or semimetallic systems such
as $\alpha$--CuMnSb, where the long-range RKKY coupling is present.

An opposite conclusion comes from the analysis of $\it single\ spin$
excitations from the AFM001 ground state. We use a $2a\times 2a\times
2a$ supercell to calculate the energy differences $\Delta E$ for  the
following cases, in which we change
(i) spin of one Mn ion, $1\mathrm{Mn}\!\uparrow\to
1\mathrm{Mn}\!\downarrow$, called a single spin-flip,
(ii)  $2\mathrm{Mn}\!\uparrow\to 2\mathrm{Mn}\!\downarrow$ for
spins of two nearest Mn ions belonging to one layer and
(iii) $2\mathrm{Mn}\!\uparrow\to 2\mathrm{Mn}\!\downarrow$ for two
distant Mn ions. 
In these processes the long-range coupling is not important, and indeed
the calculated exchange constant consistently is 
$J_{sr}\approx -0.4$~meV.

\subsection{Crystal and magnetic properties of $\beta$--CuMnSb}
\label{sec:beta}

We now consider two possible structures of the secondary phase 
proposed based on the experimental results. 
They are characterized by doubling the periodicity in 
the [001] direction. The unit cell of $\beta$--CuMnSb, shown
in Fig.~\ref{fig:str_PD}, is tetragonally deformed relative to that of
$\alpha$--CuMnSb, with the corresponding lattice parameters
$a=5.88$~\AA\ and $c=6.275$~\AA. They differ by about 3 per cent from
our calculated cubic $a$($\alpha$--CuMnSb)~$= 6.105$~\AA. The two
interlayer spacings between the consecutive MnSb planes in the [001]
direction in the unit cell, shown in Fig.~\ref{fig:str_PD}~(b), are 
quite
different, namely $d_{inter 1} = 2.80$~\AA\ (no Cu), and $d_{inter 2} =
3.48$~\AA\ (with Cu). Turing to the magnetic order of $\beta$--CuMnSb,
we find that the FM phase constitutes the ground state with 
$m_\mathrm{sat}=4.6\mu_\mathrm{B}$  
and is lower
than the AFM phase by 11~meV per f.u., as indicated in 
Tab.~\ref{tab:mag}. Thus, the
two crystalline phases of CuMnSb are in two different magnetic phases.

The experimental~\cite{Malavi} lattice parameters of $\beta$--CuMnSb 
reasonably agree with our values, i.e., the calculated $a=6.28$~\AA\ and 
$c/a=1.87$ are about 2\% larger than those measured for the 
$compressed$ crystal at the critical pressure of 7~GPa. 
On the other hand, the calculations of Ref.~\onlinecite{Malavi} predict 
that 
the 
magnetic order of the $\beta$~phase is AFM, in striking contrast with our 
results. Also their calculated $m_\mathrm{sat}(Mn)=3.8 \mu_\mathrm{B}$ 
is substantially smaller than our $4.6 \mu_\mathrm{B}$. The origin of 
these discrepancies is not clear, but it may be due to the different  
exchange-correlation functionals used, and/or to application of the 
$+U$(Mn) correction in our calculations (which can affect the 
results.~\cite{Pickett})

\begin{figure*}[t!]
\includegraphics[width=0.45\textwidth]{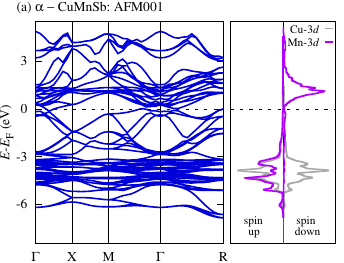}
\includegraphics[width=0.45\textwidth]{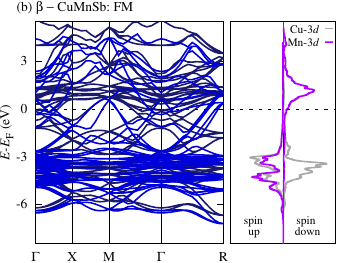}
\caption{Bands and partial DOSs for (a) the AFM001 of $\alpha$--CuMnSb 
and for
(b) the FM state of $\beta$--CuMnSb obtained using the
$a\times a\times c$ cell. The right panels show the partial DOSs for Cu
and Mn ions, and thus different contributions to spin-up and spin-down
density of Mn$\uparrow$ are exposed also in AFM case. In (b) the spin
degeneracy is lifted.
}
\label{fig:dos}
\end{figure*}

The calculated total energy of the FM $\beta$--CuMnSb relative to the
AFM $\alpha$--CuMnSb is higher by 102~meV per f.u. This energy difference 
is
not large, being comparable to the growth temperature, which implies that
the $\beta$--CuMnSb polymorph can indeed form during epitaxy. We also
stress that stoichiometry of the $\alpha$ and $\beta$~phases is the same,
which facilitates formation of $\beta$--CuMnSb. Finally, the observed
$\beta$--CuMnSb inclusions are coherent, i.e., lattice matched, with the
host structure. This agrees with the fact that the calculated excess 
elastic energy of matching the lattice parameters of the $\beta$~phase 
to the host $\alpha$~phase is very low and ranges from 
3~meV per f.u. (when the tetragonal $a$
parameter constrained to the cubic $a=6.105$~\AA) to 20~meV per f.u. (the
tetragonal $c$ parameter constrained to the cubic $a$).

The second considered possibility, Cu$_3$Mn$_2$Sb$_2$ shown in
Fig.~\ref{fig:str_PD}~(c), is higher in energy by 0.37~eV per f.u. 
in the Cu--rich conditions than the ideal CuMnSb, i.e., by 0.27~eV per 
f.u. than $\beta$--CuMnSb, its stoichiometry is markedly different, 
and thus we can eliminate this structure from  considerations.


\subsection{Energy band structures of $\alpha$--$\mathbf{CuMnSb}$ and
$\beta$--$\mathbf{CuMnSb}$}
\label{sec:bands}

{Figure~\ref{fig:dos}}~(a) shows the energy bands and DOS of the
AFM001 $\alpha$--CuMnSb. We see that this phase has a metallic
character, however DOS at the Fermi level is low. The states close to
$E_\mathrm{F}$ are built from $s$, $p$ and $d$ states of all ions with
similar weights. The low DOS($E_\mathrm{F}$) makes CuMnSb almost
semimetallic with a low electrical conductivity. Compatible with the 
small
DOS$(E_\mathrm{F})$ is the high resistivity measured in
Ref.~\onlinecite{Regnat,comment2}.

Since the system is antiferromagnetically ordered, the total DOSs 
of spin-up and spin-down states are the same. In Fig.~\ref{fig:dos} 
only contributions
of the $3d$(Mn) and $3d$(Cu) orbitals are presented to reveal
magnetic properties. We see that the exchange spin splitting of the
$d$(Mn) shell is large, about 5~eV. The closely spaced levels 
contributing
to the DOS maxima centered at 4~eV below the Fermi energy are
composed mainly of the $d$ states of both Cu and Mn. Spin-up and 
spin-down $3d$(Cu) orbitals are almost completely occupied, 
and thus Cu ions
are non-magnetic. In turn, the majority spin states of the $3d$(Mn) 
orbitals
are completely occupied, while the minority spin states at 1~eV above the
Fermi energy are partially filled thanks to a small overlap with spin up
states. As a result, a single Mn ion is in between the $d^5$ and $d^6$
configuration, with the saturation magnetic moment of $4.6
\mu_\mathrm{B}$ consistent with Tab.~\ref{tab:mag}. Our results for
$\alpha$--CuMnSb are close to those of Ref.~\onlinecite{Pickett}. A 
similar
electronic configuration takes place in CuMnAs, 
where the spin-down Mn states are partially filled.~\cite{Maca-CuMnAs}

The overall band structure of the FM $\beta$--CuMnSb displayed in
Fig.~\ref{fig:dos}~(b) is close to that of $\alpha$--CuMnSb, which is
particularly clear when comparing partial DOS of both phases.
In particular, $m_\mathrm{sat}$(Mn) is about $4.5 \mu_\mathrm{B}$ in 
both phases,
and energies of both $d$(Mn)- and $d$(Cu)-related bands are largely
independent of the actual crystal structure. This similarity can be due 
to
the fact that the MnSb (001) planes play a dominant role, and the exact
locations of the Cu ions are less important.

On the other hand, the calculated DOS($E_\mathrm{F}$) for the 
$\alpha$~phase is 0.35~states per spin and f.u., 
while for the $\beta$~phase we find 
1.26~states per spin and f.u., which is 3.6 times higher. As a 
consequence,
$\alpha$--CuMnSb is semimetallic, and the AFM order is dominant, while
$\beta$~phase is more metallic in character, which in turn favors the 
RKKY-type coupling and the FM order. 
This feature can explain the different
magnetic phases of the $\alpha$ and $\beta$ polymorphs.

Analysis of the calculated electronic structure of Heusler and 
half-Heusler
CuMnZ led Sasioglu {\it et al.}~\cite{Sasioglu} to the conclusion that 
when the spin polarization of conduction electrons is large, 
and the $d$(Mn) spin
down states are far above $E_\mathrm{F}$, then the RKKY coupling is
dominant, and one should expect the FM order, otherwise the short range
AFM coupling is dominant. Our results do not confirm this conclusion, and
indicate that the important factor determining the magnetic order is the
DOS($E_\mathrm{F}$).

\subsection{Point native defects in $\alpha$--CuMnSb}
\label{sec:eform}

Formation energy of a defect D is given by
\begin{equation}
E_{form}(\mathrm{D})\! =\! E(\mathrm{CuMnSb\!:\!D})\! -\!
E(\mathrm{CuMnSb})\!
+\! \sum_i \! {n_i \mu_i},
\label{eq:eform}
\end{equation}		
where $E(\mathrm{CuMnSb})$ and $E(\mathrm{CuMnSb:D})$ are the total
energies of a supercell without and with a defect, and $n_i = +1 (-1)$
corresponds to the removal (addition) of one $i$th atom. $\mu_i$s are the
variable chemical potentials of atoms in the solid, which in general are
different from the chemical potentials $\mu_i(\mathrm{bulk})$ of the
standard state of elements, i.e., Cu, Mn and Sb bulk. 
Details of calculations of chemical potentials are given 
in~Appendix~\ref{Appendix}.

\begin{figure}[t]
\includegraphics[width=0.5\textwidth]{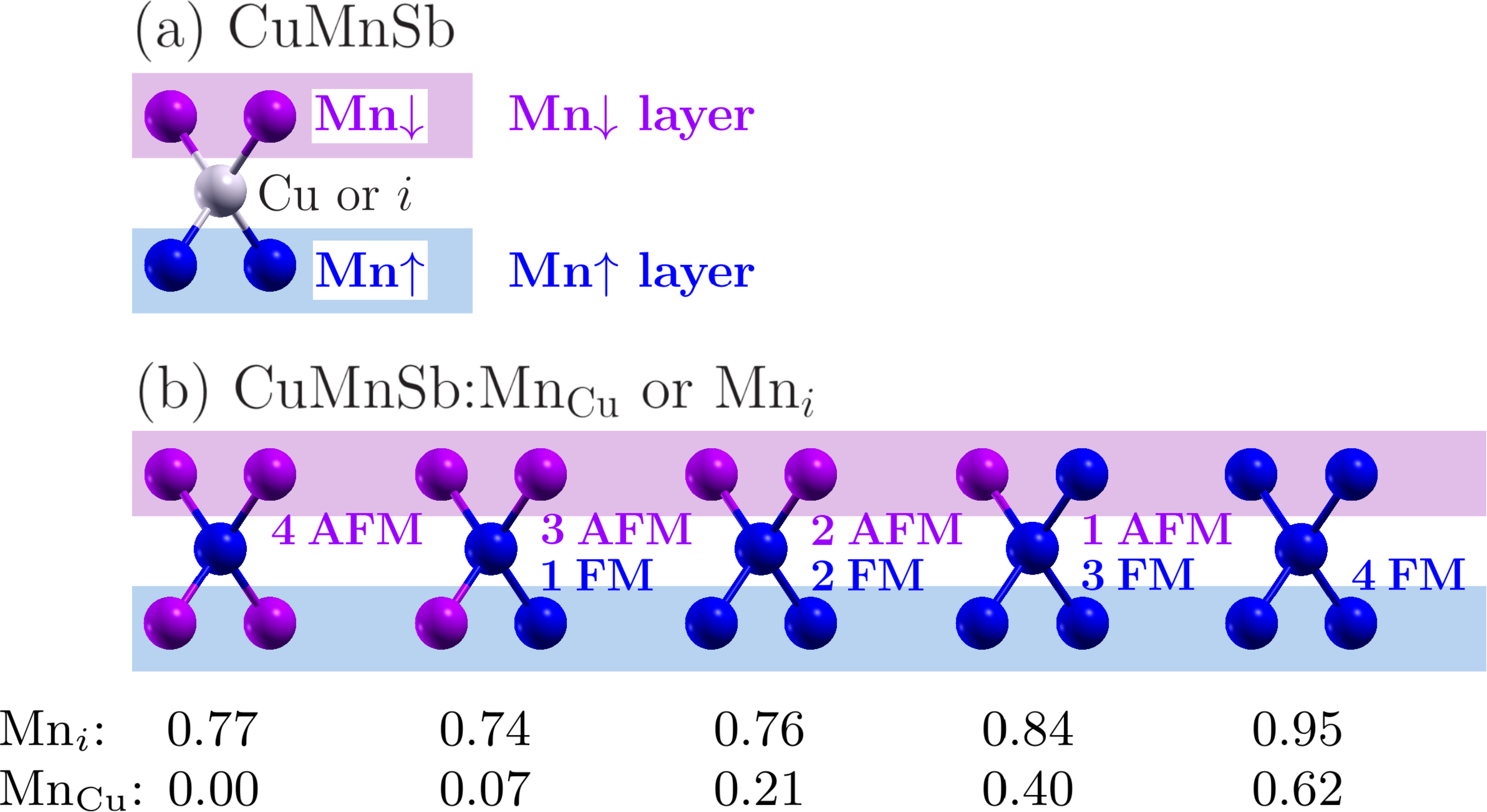}\\
\caption{All possible spin orientations of Mn ions in $\alpha$--CuMnSb
AFM001.
In the ground state configuration, the Mn spins are parallel within each
MnSb (001) layer, and the consecutive MnSb (001) layers are AFM, as shown
in (a). (b) Mn $\mathrm{Mn}_\mathrm{Cu}$ antisites and 
$\mathrm{Mn}_i$ interstitials can
assume 5 different local spin configurations. 
The corresponding spin-dependent formation energies in eV are given by 
the numbers below. }
\label{fig:Mn_spin}
\end{figure}

The point native defects considered here are vacancies $V_X$, 
interstitials 
X$_i$, and antisites X$_Y$ (where X and Y are Cu, Mn, or Sb) for all 
three 
sublattices. As it was mentioned above, the Cu sublattice  is "half- 
empty" 
compared to the MnSb sublattice. Consequently, we  consider here 
formation of interstitials at the empty sites of the Cu  sublattice only, 
and 
neglect other possibilities, expected to have higher formation 
energies $E_{form}$. 
Thus, the set of defects considered here only partially overlaps with 
that of 
Ref.~\onlinecite{Maca94}. Of particular interest to the  present study 
are 
defects 
involving Mn ions, since they can influence  magnetic properties of 
$\alpha$--CuMnSb~.\cite{Maca94} This is why we consider them more 
extensively, after briefly analyzing the non-magnetic defects. The 
calculated formation energies are summarized in  
{Tab.~\ref{tab:eform1}}. Because of 
the magnetic coupling, formation energies of the Mn-related defects 
depend on the spin direction relative to the spins of the host Mn 
neighbors. We consider possible spin configurations shown in 
{Fig.~\ref{fig:Mn_spin}} (b). 

\begin{table*}[t!]
\caption{Formation energies (in eV) of isolated point defects in the
Mn-rich conditions. In parentheses are Mn-related values corrected
for $\Delta H_f(\mathrm{MnSb})=0.48$~eV,
which correspond to the Mn-poor case.}
\label{tab:eform1}
\begin{ruledtabular}
\begin{tabular}{p{0.12\textwidth}p{0.1\textwidth}|p{0.12\textwidth}p{0.1\textwidth}|p{0.12\textwidth}p{0.23\textwidth}}
Cu defects && Sb defects && Mn defects\\
\hline
$V_{\mathrm Cu}$     		& -0.01       & $V_\mathrm{Sb}$     	
	& 2.45 & $V_\mathrm{Mn}$ & -0.07 (-0.55)\\
$\mathrm{Cu}_i$     		& 1.00        & $\mathrm{Sb}_i$     	
	& 3.64 & $\mathrm{Mn}_i$ & 0.74--0.95 (1.22--1.43)\\
$\mathrm{Cu}_{\mathrm Mn}$   & 0.48 (0.00) & $\mathrm{Sb}_\mathrm{Mn}$   
&
1.60 (1.12) & $\mathrm{Mn}_\mathrm{Cu}$   & 0.00--0.62 (0.48--1.10) \\
$\mathrm{Cu}_\mathrm{Sb}$   & 1.20        & $\mathrm{Sb}_\mathrm{Cu}$   &
2.82 & $\mathrm{Mn}_\mathrm{Sb}$   & 1.18 (1.65) \\
\end{tabular}
\end{ruledtabular}
\end{table*}

Formation energy determines the corresponding equilibrium 
concentration [$D$] of a defect $D$ according to
\begin{equation}
[D]=N_0 \exp[-E_{form}(D)/k_\mathrm{B} T],
\end{equation}

\noindent
where $k_\mathrm{B}$ is the Boltzmann constant 
and $N_0$ is the density of the relevant
lattice sites. 
Details of the calculations of $E_{form}$ are provided in Supporting  
Information. 
To put the calculated formation energies into a proper context, we note
that if the growth temperatrure 
$T_{growth}=250^0$C and $E_{form}=0.1$~eV, then  
$\exp(-E_{form}/k_\mathrm{B} T_{growth}) = 0.1$, which corresponds to a 
high 10 atomic 
per cent concentration of this defect on the considered sublattice. 
On the other hand, if $E_{form}=1$~eV, then  
$\exp(-E_{form}/k_\mathrm{B} T_{growth}) = 9\times 10^{-11}$, 
which implies a negligible defect concentration.

{\bf Sb sublattice.}  The prohibitively high values of
$E_{form}$ demonstrate that $V_\mathrm{Sb}$ and $\mathrm{Sb}_i$ should
not form. Similarly, formation energies of $\mathrm{Sb}_{\mathrm Cu}$,
$\mathrm{Sb}_\mathrm{Mn}$, $\mathrm{Cu}_\mathrm{Sb}$ and
$\mathrm{Mn}_\mathrm{Sb}$ antisites exceed 1~eV, and those defects are
not expected to be present at high concentrations. Consequently, the Sb
sublattice is thermodynamically stable, robust, and constitutes
a defect-free back-bone of CuMnSb.

{\bf Cu and Mn sublattices.}
The properties of both Cu and Mn sublattices are opposite to those of the 
Sb sublattice, as they are susceptible to contain defects.  In 
particular, 

(i) Both the Cu and Mn vacancies can be present at high concentrations, 
since their $E_{form}$ are low.   

(ii) Formation energy of Cu interstitials at the Cu sublattice, 
$E_{form}(\mathrm{Cu}_i)=1$~eV, is relatively high, and their 
concentrations are negligible. Additionally, the high formation energy of 
$\mathrm{Cu}_i$ interstitials is consistent with the sparse character of 
the 
Cu sublattice in $\alpha$--CuMnSb.  

(iii) Formation of $\mathrm{Mn}_i$ interstitials at the Cu sublattice is 
characterized by $E_{form}$ = 0.7-1.4~eV, depending on the spin direction 
and conditions of growth, and therefore they are not expected to be 
present at high concentrations, especially in the Mn-poor conditions.  

(iv) Finally, $\mathrm{Mn}_\mathrm{Cu}$ antisites, with formation 
energies 
ranging from 0 to about 1~eV, can be present at high concentrations, 
comparable to those of $V_\mathrm{Mn}$ and $V_\mathrm{Cu}$. 

In brief, low formation energies are found for three defects, namely 
the $V_\mathrm{Cu}$ and $V_\mathrm{Mn}$ vacancies and the 
$\mathrm{Mn}_\mathrm{Cu}$ antisite, particularly at the Mn-rich growth 
conditions. This indicates that a Cu deficit on the Cu sublattice is 
possible, 
affecting stoichiometry. Significantly, $\mathrm{Mn}_\mathrm{Cu}$ 
antisites 
make the Cu sublattice magnetic, and also they can participate in the 
magnetic coupling between the adjacent MnSb (001) planes, thus 
influencing magnetic properties, as it will be discussed in more detail 
below. In contrast, $\mathrm{Sb}_\mathrm{Cu}$ antisites are present in 
negligible concentrations. Our results are in a reasonable agreement with 
those of Ref.~\onlinecite{Maca94}, especially given their neglect of spin 
effects 
and a somewhat different theoretical approach. Interestingly, formation 
energies of native defects in CuMnAs calculated in 
Ref.~\onlinecite{Maca-CuMnAs}  
are close to the present results in spite of the different anion.

\subsection{Defect-induced magnetic coupling}
\label{sec:Mn_defects}

There are two Mn-related point defects, $\mathrm{Mn}_i$ and 
$\mathrm{Mn}_\mathrm{Cu}$, both situated on the Cu sublattice. When  
present at high concentrations, they affect magnetism of 
$\alpha$--CuMnSb. Their  coupling with host Mn ions is different than the 
Mn-Mn 
coupling between  the  host Mn because of the different local 
coordination. Energetics of both  defects is complex and rich, since the 
total energy of the system (and thus formation energies) depends on their 
spin orientations relative to the neighborhood. At both substitutional 
and 
interstitial sites in the Cu layer, a Mn ion  has 4 Mn nearest neighbors 
arranged in a tetrahedral configuration, 2 in the upper and 2 in the 
lower 
MnSb  layer.  The $\mathrm{Mn}_i$--$\mathrm{Mn}_\mathrm{Mn}$  distance 
is shorter than that  of 
$\mathrm{Mn}_\mathrm{Mn}$--$\mathrm{Mn}_\mathrm{Mn}$, and equal to  
$(\sqrt{3}/4) a$.

The possible local spin configurations are reduced to small  clusters of 
5 Mn 
ions, shown in Fig.~\ref{fig:Mn_spin}. The Mn spin-up  and spin-down  
(001) MnSb layers are denoted by in pink and blue, respectively,  
reflecting 
the calculated (001) AFM magnetic ground state. The  central 
$\mathrm{Mn}_\mathrm{Cu}$  (or $\mathrm{Mn}_i$) ion of such  a cluster 
provides an additional channel of magnetic coupling  between two 
adjacent MnSb layers. The corresponding formation energies are given in 
Fig.~\ref{fig:Mn_spin}.

As it was pointed out, in ideal  $\alpha$--CuMnSb, the Mn ions are second 
neighbors only, separated  either by Sb (i.e., the Mn-Sb-Mn "bridge" in 
the 
MnSb(001) plane), or  by Cu (forming a Mn-Cu-Mn "bridge" linking 3 
consecutive (001)  planes.) Thus, the short range magnetic coupling in 
ideal 
$\alpha$--CuMnSb is successfully modelled  in Sec.~\ref{sec:alpha} by the 
interaction between two Mn {\it  second neighbors}, situated either in 
the 
same MnSb layer, or in two  adjacent ones. In contrast, the 4 host Mn 
ions 
in the cluster are the  {\it first} neighbors of a $\mathrm{Mn}_i$ or a  
$\mathrm{Mn}_\mathrm{Cu}$ defect. Thus, one can expect that this  
coupling is stronger than the intrinsic one in the ideal host, 
and indeed, the differences in energy between various configurations in 
Fig.~\ref{fig:Mn_spin} are about 100~meV, which is too high to be 
explained by the estimated $J_{sr}=0.4$~meV. 

As it follows from Fig.~\ref{fig:Mn_spin}, 5-atom clusters are 
magnetically frustrated. In particular, the lowest energy case denoted as 
4AFM favors the local FM orientation of spins in two adjacent (001) 
planes, 
which is opposite to the global host magnetic order.  Our results do not 
confirm the conclusion of Ref.~\onlinecite{Maca-CuMnAs} who find that 
the 3AFM configuration has the lowest energy, and thus it promotes the 
global AFM111 order. Instead, we rather expect that Mn-related point 
defects induce disorder of the host  AFM phase, possibly leading to 
formation of a spin glass.~\cite{Kroder}

\section{Summary}
\label{sec:compar}

CuMnSb films were epitaxially grown on GaSb substrates. Magnetic
measurements reveal the presence of two magnetic subsystems. The dominant
magnetic order is AFM with the N\'{e}el temperature of 62~K, which is the
same as in bulk CuMnSb. It co-exists with a FM phase, characterized by
the Curie temperature of about 100~K.

These findings go in hand with transmission electron microscopy and
selective area diffraction measurements, which  demonstrate coexistence
of two structural polymorphs of the same stoichiometry. The dominant one
is the cubic half-Heusler $\alpha$--CuMnSb, which is the equilibrium
structure of bulk samples. The  second component is a  tetragonal
$\beta$--CuMnSb polymorph, which forms 10-100 nm long elongated 
inclusions.

The results of our ab-initio calculations provide a consistent
interpretation of the experimental data, and in particular

(i) The $\beta$--CuMnSb phase is metastable, and its total energy is
higher by 0.1 eV per f.u. only than that of the equilibrium 
$\alpha$--CuMnSb.
Lattice parameters of the $\beta$~phase differ from those of 
$\alpha$--CuMnSb 
by about 4 per cent. This lattice misfit between the two
structures does not prevent the pseudomorphic coexistence of both
phases, since the calculated misfit strain energy is below 20~meV per 
f.u.

(ii) In agreement with experiment, $\alpha$--CuMnSb is AFM, and the FM
order is 19 meV per f.u. higher in energy. In contrast, the magnetic 
ground
state of $\beta$--CuMnSb is FM, which is more stable than AFM by 11
meV per f.u. This indicates that indeed the $\beta$--CuMnSb inclusions 
are
responsible for the FM signal.

(iii) The different magnetic orders of the $\alpha$ and $\beta$~phases
originate in their somewhat different band structures. In particular,
critical for magnetic order is the DOS at the Fermi level, which is about
4 times higher in $\beta$--CuMnSb than in the $\alpha$~phase. This shows
that the $\beta$~phase is more metallic in character, which in turn
favors the FM order driven by the Ruderman-Kittel-Kasuya-Yoshida
interaction.

(iv) Our calculations predict the saturated magnetic moment of Mn
$m_\mathrm{sat}= 4.6 \mu_\mathrm{B}$ and $4.5 \mu_\mathrm{B}$ for the
$\alpha$
and the $\beta$~phase, respectively. This corresponds to the effective
moment of $5.6\mu_\mathrm{B}$, in good agreement with the measured
$5.5\mu_\mathrm{B}$.

(v) The calculated formation energies of point native defects indicate
that the most probable are the Mn$_\mathrm{Cu}$ antisites with low 
formation
energies of 0--0.2 eV. However, their presence is expected to disorder
the host magnetic AFM phase rather than to induce a transition to the FM
configuration.

(vi) Regarding the properties of the CuMnX series we see that their
structural stability is relatively weak, as they crystallize in a variety
of structures. In particular, unlike the bulk
orthorhombic CuMnAs, epitaxial films of CuMnAs are tetragonal, but both
structures are AFM. In the case of CuMnSb, polymorphism comprises also
the equilibrium magnetic structure, AFM in the bulk specimens, and FM in
epitaxial films.

\acknowledgements 
LS, CG, JK and LWM thank M. Zipf for technical assistance. 
Our work was funded by 
the Deutsche Forschungsgemeinschaft (DFG,German Research
Foundation) No. 397861849, by the Free State of Bavaria (Institute 
for Topological Insulators) and the Deutsche Forschungsgemeinschaft 
(DFG, German Research Foundation) under Germany's Excellence 
Strategy - EXC2147 ct.qmat (Project-Id 390858490).

\appendix\section{\label{Appendix}}
The difference
between chemical potential $\mu_i$ considered in formation energy, 
eq.~\ref{eq:eform}, and the chemical potential of bulk Cu, Mn or Sb, 
$\mu_i(\textrm{bulk})$, is denoted by $\delta \mu_i$:
\begin{equation}
\mu_i = \mu_i(\textrm{bulk}) + \delta \mu_i.
\end{equation}
The highest possible value of $\mu_i$ is $\mu_i(\textrm{bulk})$, which
implies that the studied system is in equilibrium with the given bulk
source of atoms and $\delta \mu_i=0$, otherwise $\delta \mu_i <0$.

Chemical potentials of the components in the standard state are given by
the total energies per atom of elemental solids. The calculated cohesive
energies $E_{coh}$ of the face centered cubic Cu, the face centered cubic
Mn with the AFM magnetic order, and the triclinic Sb are,
respectively, 3.40
(3.49), 2.65 (2.92) and 2.68 (2.75)~eV/atom. They compare reasonably well
with the experimental values given in parentheses.~\cite{Kittel}  

Chemical potentials of the involved atomic species depend on possible
formation of compounds. The ranges of variations of chemical potentials
are determined by conditions of equilibrium between various phases, i.e.,
Cu$_2$Sb, MnSb and CuMnSb. Thermodynamic equilibrium requires that

\begin{eqnarray}
\label{eq:m}
&&\delta \mu(\textrm{Cu}) + 2 \delta \mu(\textrm{Sb}) = \Delta
H_f(\textrm{Cu}_2\textrm{Sb}),\nonumber \\
&&\delta \mu(\textrm{Mn}) + \delta \mu(\textrm{Sb}) =
\Delta H_{f}(\textrm{MnSb}), \\
&&\delta \mu(\textrm{Cu}) + \delta \mu(\textrm{Mn}) +
\delta \mu(\textrm{Sb}) = \Delta H_{f}(\textrm{CuMnSb}), \nonumber
\end{eqnarray}
where $\Delta H_{f}$ is the enthalpy of formation per formula unit
(negative for a stable compound).

The calculated values $\Delta H_{f}(\textrm{Cu}_2\textrm{Sb}) = -
0.03$~eV per f.u., $\Delta H_{f}(\textrm{MnSb}) = -0.48$~eV per f.u.,
and $\Delta H_{f}(\textrm{CuMnSb}) = -0.42$~eV per f.u. The very low 
$\Delta
H_{f}(\textrm{Cu}_2\textrm{Sb})$ is somewhat unexpected, since
$\textrm{Cu}_2\textrm{Sb}$ is a stable compound which crystallizes in the
tetragonal phase.~\cite{Endo} Next, our result $\Delta
H_{f}(\textrm{MnSb}) = -0.48$~eV per f.u. agrees well with both the 
previous
value -0.52~eV per f.u. calculated in Ref.~\onlinecite{Podloucky}, and 
the
experimental -0.52 eV per f.u.~\cite{Shchukarev} 
Assuming that the accuracy of the calculated values is 0.03~eV per f.u., 
the
set of {Equation~\ref{eq:m}} is consistent if we assume $\Delta
H_{f}(\textrm{Cu}_2\textrm{Sb}) = 0$, and $\Delta H_{f}(\textrm{MnSb}) =
\Delta H_{f}(\textrm{CuMnSb}) = -0.45$~eV per f.u.
This in turn implies that $\delta \mu(\textrm{Cu})=\delta
\mu(\textrm{Sb})=0$, and $\delta \mu(\textrm{Mn})=-0.45$~eV.
Consequently, the allowed window of the Mn chemical potential is
\begin{equation}
-0.45 ~\textrm{eV}< \delta \mu(\textrm{Mn}) < 0,
\end{equation}
where the two limiting values correspond to the Mn-poor and Mn-rich
conditions, and an analogous window holds for Sb.



\begin{thebibliography}{61}
\bibitem [{com({\natexlab{b}})}]{comment2}%
  \BibitemOpen
  \bibinfo {note} {in addition, resistivity
  is increased by native defects. Indeed, the estimated concentration of
  antisites involving Cu and Mn sublattices is 1.5 atomic \%.~\cite{Regnat,
  Maca94}}\BibitemShut {NoStop}%
\end{thebibliography}

\label{bib}

\end{document}